# Coronavirus (COVID-19) Classification using Deep Features Fusion and Ranking Technique

Umut Ozkaya[1], Saban Ozturk[2], Mucahid Barstugan[1]

Electrical and Electronics Engineering, Konya Technical University[1], Konya, Turkey
Electrical and Electronics Engineering, Amasya University[2], Amasya, Turkey
Corresponding Author: uozkaya@ktun.edu.tr

**Abstract:**
Coronavirus (COVID-19) emerged towards the end of 2019. World Health Organization (WHO) was identified it as a global epidemic. Consensus occurred in the opinion that using Computerized Tomography (CT) techniques for early diagnosis of pandemic disease gives both fast and accurate results. It was stated by expert radiologists that COVID-19 displays different behaviours in CT images. In this study, a novel method was proposed as fusing and ranking deep features to detect COVID-19 in early phase. 16x16 (Subset-1) and 32x32 (Subset-2) patches were obtained from 150 CT images to generate sub-datasets. Within the scope of the proposed method, 3000 patch images have been labelled as CoVID-19 and No finding for using in training and testing phase. Feature fusion and ranking method have been applied in order to increase the performance of the proposed method. Then, the processed data was classified with a Support Vector Machine (SVM). According to other pre-trained Convolutional Neural Network (CNN) models used in transfer learning, the proposed method shows high performance on Subset-2 with 98.27% accuracy, 98.93% sensitivity, 97.60% specificity, 97.63% precision, 98.28% F1-score and 96.54% Matthews Correlation Coefficient (MCC) metrics.

**Keywords:** Coronavirus, COVID-19, CT images, Deep Learning, Feature Fusion and Ranking.

## 1. INTRODUCTION

Corona virus disease (COVID-19) is essential to apply the necessary quarantine conditions and discover the treatment methods in order to prevent the rapid spread of COVID-19. It has become a global epidemic similar to other pandemic diseases, causes patient deaths in China according to World Health Organization (WHO) data [1-3]. Early application of treatment procedures for individuals with COVID-19 infection increases the patient's chances of survival.

Fever, cough and shortness of breath are the most important symptoms in infected individuals for the diagnosis of COVID-19. At the same time, these symptoms may show carrier characteristics by not being seen in infected individuals. Pathological tests performed in laboratories are taking more time. Also, the margin of error can be high. A fast and accurate diagnosis is necessary for an effective struggle against COVID-19. For this reason, experts have been started to use radiological imaging methods. These procedures are performed with computed tomography (CT) or X-ray imaging techniques. COVID-19 cases have similar features in CT images in the early and late stages. It shows a circular and inward diffusion from within the image [4]. Therefore, radiological imaging provides early detection of suspicious cases with an accuracy of 90%.

When the studies in the literature are examined, Shan et al proposed a neural network model called VB-Net in order to segment the COVID-19 regions in CT images. This proposed method has been tested in 300 new cases. A recommendation system has been used to make it easier for radiologists to mark infected areas within CT images [5]. Xu et al. analyzed CT images to determine healthy, COVID-19 and other viral case. The dataset used included 219 COVID-19, 224 viral diseases and 175 healthy images. They achieved 87.6% general classification accuracy with their deep learning method [6]. Apostolopoulos et al. proposed a transfer learning methods to classify COVID-19 and normal case. They obtained performance metrics which are 96.78% accuracy, 98.66% sensitivity, and 96.46% specificity [7]. Shuai et al. were able to successfully diagnose COVID-19 using deep learning models that could obtain graphical features in CT images [8].

150 CT images were used in this study to classify COVID-19 cases. Two different datasets were generated from 150 CT images. These datasets include 16×16 and 32×32 patch images. Each dataset contains 3000 number of images labeled with COVID-19 and No findings. Deep features were obtained with pre-trained Convolutional Neural Network (CNN) models. These deep features was fused and rank to train Support Vector Machine (SVM). The performance of proposed method can be used for early diagnosis of COVID-19 cases.

This study consists of 5 sections. The properties of obtained patch images are visualized in Section 2. In Section 3, the basics of deep learning methods, feature fusion and ranking techniques are mentioned. Comparative classification performances are given in Section 4. There is a discussion and conclusion in Section 5.

## 2. MATERIAL

### 2.1. Statistical Features of Dataset Used

53 infected CT images was accessed to the Societa Italiana di Radiologia Medica e Interventistica to generate datasets [9]. Patch images obtained from infected and non-infected regions form CT images. Properties of two different patch are given in Table 1.

**Table 1.** Properties of Two Different Patch Datasets

| Subset | Patch Dimension | Number of COVID-19 Patches | Number of No findings Patches |
|---|---|---|---|
| Subset 1 | 16x16 | 3000 | 3000 |
| Subset 2 | 32x32 | 3000 | 3000 |

### 2.2. Visual Features of Dataset

Different CT tools were used to generate the datasets. There are COVID-19 regions with different grey levels in the infected CT images. CT images were obtained with different devices. Thus, it caused different infected grey level region in the CT images. This situation affects the classification process and performance quite negatively. CT images have different grey levels. Image patches of $16 \times 16$ and $32 \times 32$ dimensions with different characteristics from these CT images were obtained. The process of obtaining patches is given in Figure 1.

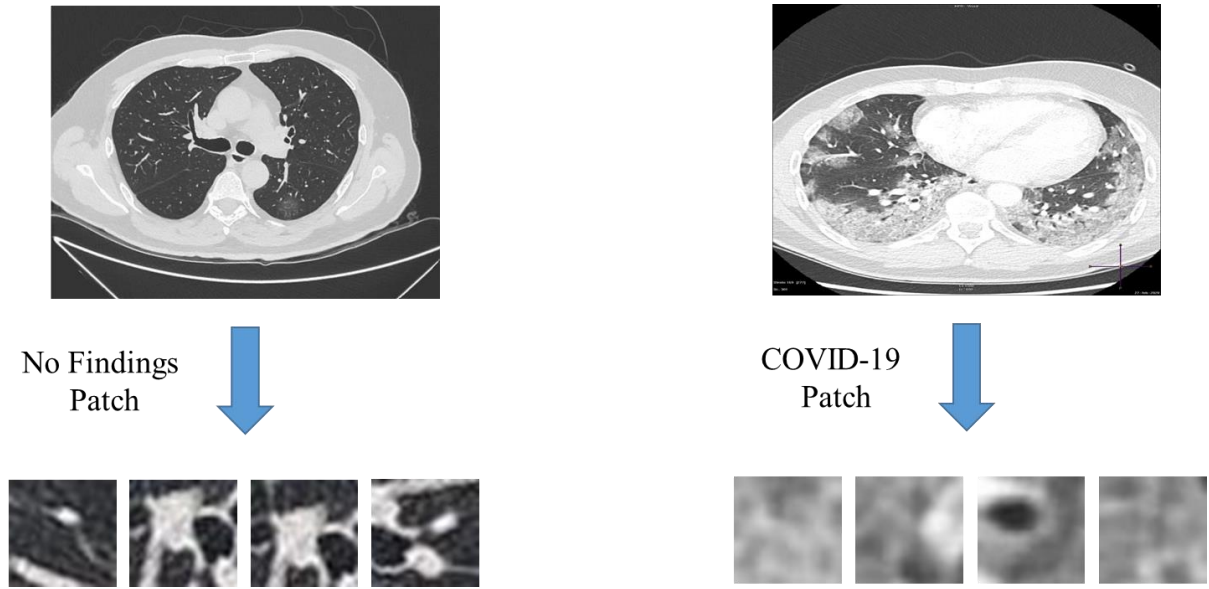

**Figure 1.** Data Generation for COVID-19 and No Findings Patches

## 3. METHOD

### 3.1. Deep Learning

In 2006, Geoffrey Hinton has shown that deep neural networks can be effectively trained by the greedy-layered pre-training method [10]. Other research groups used the same strategy to train many other deep networks. The use of term (Deep Learning) in order to draw attention to the theoretical importance of depths has been popularized for the design of better performing networks of neural networks and the importance of deeper networks.

Deep learning, which has become quite popular recently, has been used in many areas. E-mail filtering, search engine matching, smartphones, social media, e-commerce can be written to them. Academic studies have been pioneers for their use in these areas. Deep learning is also used for face recognition, object recognition, object detection, text classification and speech recognition. Deep learning is a type of artificial neural network and has multilayers. The more layers are increased, the greater accuracy is achieved. While deep convolutional networks are successfully used in image, video, speech and sound processing, recurrent neural networks are used in sequential data such as text and speech.

Deep learning, started to be used in 2010, a large data set with multilayer of machine learning calculations used in many layers, even in the machine learning the parameters that need to be defined, perhaps a better system that can evaluate the parameters. Deep learning artificial neural networks are the algorithms created by taking advantage of the functions of the brain.

In machine learning, Deep Belief Networks (DBN) is a productive graphical model or, alternatively, a class of deep neural networks consisting of multiple layers in hidden nodes. When trained on a series of unsupervised examples, the DBN can learn to reconfigure its entries as probabilistic. The layers then act as feature detectors. After this learning phase, a DBN can be trained with more control to make the classification. DBNs can be seen as a combination of simple, unsupervised networks, such as restricted Boltzmann machines (RBMs) or auto encoder, which serve as the hidden layer of each subnet, the visible layer of the next layer.

*3.2. Convolutional Neural Network*

Convolution is used as a mathematical process. It is a special type of linear operations. Convolutional neural networks (CNN) are a type of neural network with at least one layer of convolution. However, the convolution process in deep learning is different from the convolution process in normal or engineering mathematics. Convolution neural networks has some layer such as convolution, ReLU, Pooling, normalization, fully connected and softmax layer. In the convolution neural networks, classification process takes place in fully connected layers and softmax layer.

Generally, convolution is a process that takes place on two actual functions. To describe the convolution operation, two function can be used for this definition. For example, the location of a space shuttle with a laser is monitored. The laser sensor produces a simple x(t) output, which is the space of the space shuttle at time t. Where X and t are actual values, for example, any t is a different value received at a snapshot time. Also, this sensor has a bit noisy. To carry out a less noisy prediction, designer can take the average of several measurements together. Naturally, final measurements are closer, so that the average weights that give more weights to desired final measurements. This can be done with the weighting function w(a), which is a measurement period. If a weighted average operation is applied at all times, a new function is obtained which allows to more accurately estimate the position:

$$S(t) = \int X(a)W(t-a)da \qquad (1)$$

The above process is a convolution and is represented by a star:

$$S(t) = (X * w)(t) \qquad (2)$$

In CNN terminology, first argument in X function at Eq. 2 is called an introduction to convolution and the second argument for W function is called the kernel. The output is called feature map. In the above example, the measurement is made without interruption, but this is not realistic. Time is parsed when working on the computer. In order to realize realistic measurement, one measurement per second is taken. Where t is the time index and is an integer, so X and W are integers.

$$S(t) = (X*W)(t) = \sum_{a=-\infty}^{\infty} X(a)W(t-a) \qquad (3)$$

In machine learning applications, the input function consists of a multidimensional array set and the kernel function consists of a multidimensional array of several parameters. Multiple axes are convolved at one time. So if the input is a two-dimensional image, the kernel becomes a two-dimensional matrix.

$$S(i,j) = (I*K)(i,j) = \sum_m \sum_n I(i-m, j-n)K(m,n) \qquad (4)$$

The above equation means shifting the kernel according to the input. This increases invariance of convolution [11]. But this feature is not very important for machine learning libraries. Instead, many machine learning libraries process the kernel without inversion, which is called as cross correlation, which is related to convolution. But because it looks like a convolution, it is called a convulsive neural network:

$$S(i, j) = (I * K)(i, j) = \sum_m \sum_n I(i + m, j + n) K(m, n) \tag{5}$$

Discrete convolution is seen as a matrix product. Typical convolution neural networks' benefit from further expertise to effectively deals with large inputs. Figure 2 shows how the process occurs in convolution neural networks:

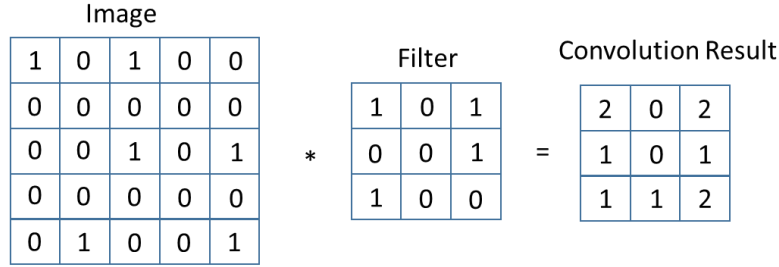

**Figure 2.** Convolution Process

Convolution provides three important thoughts to improve a machine learning system: infrequent interactions, parameter sharing, and covariant representations. Furthermore, convolution process can be worked with variable-sized inputs. Convolution neural network layers use a matrix parameter with a matrix parameter that includes a different kinds of link between each input unit and each output unit. It means that each output unit connects with each input unit. However, CNN typically have infrequent interactions (also called sparse links or sparse weights). This is done by making the kennel smaller than the entrance. Since the number of pixels after each convolution process decreases, if there is a quality that should not be overlooked at the edges, zero and edge attributes are preserved by adding zero at the end of the rows and columns. This process is called padding.

For example, input image may consist of thousands or millions of pixels for image process, but small and meaningful properties such as kernel's edges consisting of only ten or hundreds of pixels can be detected. This means we need to save fewer parameters that both reduce the memory requirements of CNN model and increase its efficiency. It also means that calculating output requires less processing. These improvements in productivity are generally quite large.

Parameter sharing refers to the use of the same parameter for more than one function in a model. In a conventional neural network, each element in weighted matrix is used to calculate the output of a layer. This is multiplied by an element of the entry and will not be reviewed again. It can be said that a network ties weights because the value of the weight applied to an input depends on the value of the weight applied elsewhere as in parameter sharing. In a CNN, each member of the core is used in each position of the insert.

Parameter sharing used by the convolution process means that instead of learning a separate set of parameters for each subject, only one set will be learned.

Each layer of a ConvNet converts one volume activation to another by a differentiable function. Three main layer types are used to create ConvNet architectures. Convolutional Layer, Pool Layer and Fully Connected Layer (exactly as seen in Normal Neural Networks). These layers are collected to create a complete ConvNet architecture. Considering that the images are three-dimensional in the form of H x W x D size if K x K is called kernel size is how many pixels of convolution output is calculated as follows:

$$\frac{H + 2P - K}{S} + 1 \tag{6}$$

Roughly means normalization. The size of the data in artificial neural networks is important. As the data grows, the memory they occupy increases and this reduces both the efficiency of the artificial neural network and decreases the working speed. By compressing the entire dataset value to 0-1, the operations are made easy. It extracts this process from the average of all the data sets and thus the data is in the range 0-1. The result of standardization) is to rescale features for a standard normal distribution.

$$\mu = 0, \sigma = 1 \tag{7}$$

where µ and σ is represented as average standard deviation respectively. Standard scores for each samples are computed as follows:

$$Z = \frac{X - \mu}{\sigma} \tag{8}$$

The standard deviation for the features is centered between 1 and 0. Also, it is important for training of many machine learning algorithms.

A pooling function changes the output of the network at a specific location with a summary statistics of nearby outputs. For example, max-pooling yields the largest in the quadrilateral space as output. Other popular pooling functions; mean and minimum pooling functions.

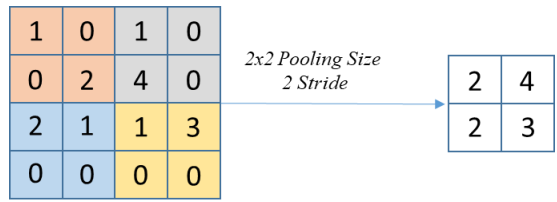

a) Maximum Pooling

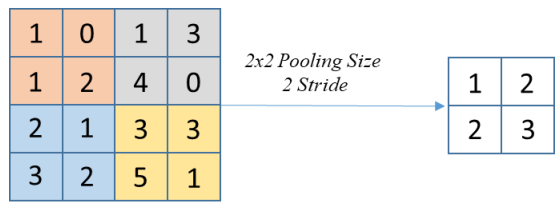

b) Mean Pooling

**Figure 3.** a) Maximum Pooling b) Average Pooling

Figure 3 shows 2 × 2 maximum pooling and mean pooling operations. The pooling process consists of outputs from certain regions. For example, when a face image is selected. The eyes are the eyes, nose and lips. The most obvious is the eyes. The attributes are different from the other parts of the face. The pooling process is used to select the attributes here. The size of the matrix is reduced during this process.

When number of parameters in the next layer depends on input image or feature map size, any reduction in input size also increases the statistical efficiency and reduces the memory requirements for storing parameters. The number of pixels of the pooling output is calculated as follows:

$$\frac{H+2P-T}{S}+1 \tag{9}$$

Rectified Linear Unit is an activation function type. The Rectified Linear Unit has recently become popular. Calculates the function F (x) = max (0, x). In other words, activation is thresholded equal to zero. There are a number of pros and cons of the use of ReLU.

It has been found that stochastic gradient descent significantly accelerates convergence compared to Sigmoid / tanh functions. It is claimed that this originates from a linear, unsatisfactory form. When the neurons containing costly operations are compared to tanh / sigmoid, ReLU can simply be applied by thresholding an activation matrix to zero.

ReLU units can become sensitive during training phase. For example, a large gradient scale flowing through neuron with a ReLU activation function can cause weights to be updated so that the neuron is not reactivated at any data point. If this happens, the gradient flowing through the unit will be zero from that point forever. That is, ReLU can kill units irrevocably during training because data replication can be disabled. For example, if the learning rate is too high, 40% of the network may be dead. This is a less frequent occurrence with an appropriate adjustment of the learning rate.

In fully connected layers, reduction of nodes below a certain threshold increased the performance. So it is observed that forgetting the weak information increases learning. Some properties of dropout value are as follows. The dropout value is generally 0.5. Different uses are also common. It varies according to the problem and data set. The random elimination method can also be used for the dropout. The dropout value is defined as a value in the range [0, 1] when used as the threshold value. It is not necessary to use the same dropout value on all layers; different dilution values can also be used.

The softmax function is a sort of classifier. Logistic regression is a classifier of the classifier and the softmax function is multi-class of logistic regression. $1/\sum_j e^{fj}$ term normalizes the distribution. That is, the sum of the values equals 1. Therefore, it calculates the probability of the class to which the class belongs. When a test input is given x, the activation function in j = 1,…,k is asked to predict the probability of p (y = j | x) for each value. For example, it is desirable to estimate the probability that the class tag will have each of the different possible values. Thus, as a result of the activation function, it produces a k-dimensional vector which gives us our predictive possibilities.

The error value must be calculated for the learning to occur and the error value for the softmax function is calculated by the softmax loss function. In the Softmax classifier, the f (xi; W) = Wxi function match remains unchanged, but we now interpret these scores as normalized log probabilities for each class and use the following form of cross entropy loss.

$$L_i = -\log\left(\frac{e^{f_{yi}}}{\sum_j e^{f_j}}\right) \quad \text{or} \quad L_i = -f_{yi} + \log\left(\sum_j e^{f_j}\right) \tag{9}$$

### 3.3. Feature Fusion and Ranking Technique

VGG-16, GoogleNet and ResNet-50 models were used for feature extraction. The obtained feature vectors with these models were fused to obtain higher dimensional fusion features. In this way, the effect of insufficient features obtained from a single CNN network is minimized. In addition, there is a certain level of correlation and excessive information among the features. This also increases consuming time and computational complexity. Therefore, it is necessary to rank the features. t-test technique was used in feature ranking. It calculates the difference between the two features and determines its differences statistically [12]. In this way, it performs the ranking process by taking into account the frequency of the same features in the feature vector and the frequency of finding the average feature.

### 3.4. Support Vector Machines (SVMs)

After the feature fusion and ranking functions were performed, the binary SVM classifier was trained for classification. SVM transfers features into space where it can better classify features with kernel functions [13]. Linear kernel function was used in SVM. The SVM classifier was trained to minimize the squared hinge loss. The squared hinge loss is given in Eq. 10.

$$\min_{w} \frac{1}{2} w^T w + C \sum_{n=1}^{N} \max(1 - w^T x_n t_n, 0)^2 \qquad (10)$$

Here, $x_n$ represents the fusion and the ranking feature vector. The wrong classification penalty is determined by the C hyper parameter in the loss function.

### 3.5. Proposed Method

In the proposed method, pre-trained CNN networks were trained for Subset-1 and Subset-2 separately. VGG-16, GoogleNet and ResNet-50 models were used as a pre-trained network. Patch images were given as input to trained pre-trained CNN structures during the test phase. Feature vectors (1000 × 1 × 3) obtained from these networks provide a new feature set with fusion process. Correlation values between features were taken into consideration in fusion process. The obtained features were ranked by t-test method. In the t-test ranking process, features close to each other were eliminated according to feature frequency. In the last stage, fusion and ranking deep features were evaluated with SVM classifier. The method proposed in Figure 4 is visualized.

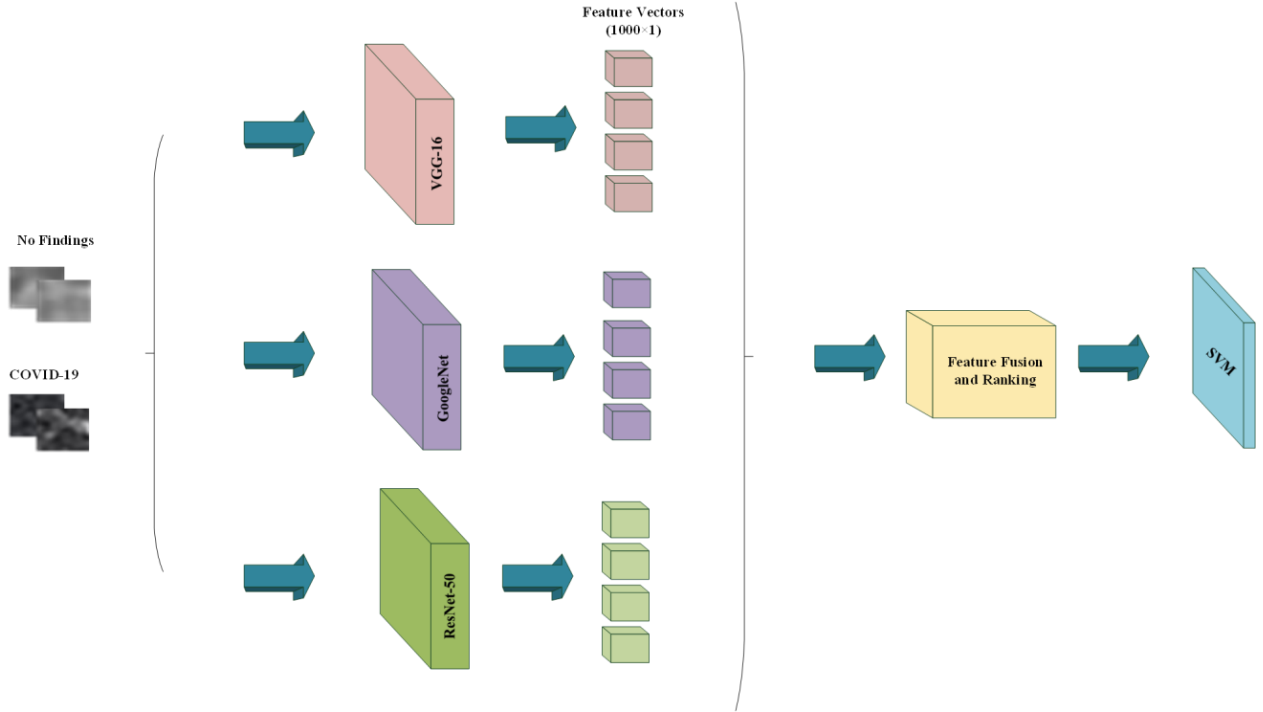

**Figure 4.** Proposed Method

## 4. EXPERIMENTAL RESULTS

This study presents the classification of COVID-19 texture for two different size data sets. First of all, the features obtained from CNN structures have been passed through fusion and ranking processes and classified. Six different metrics were used to evaluate the proposed method. These metrics are sensitivity, specificity, accuracy, precision, F-score and Matthews Correlation Coefficient (MCC).

$$Sensitivity = TP/(TP+FN) \tag{11}$$

$$Specificity = TN/(TN+FP) \tag{12}$$

$$Accuracy = (TP+TN)/(TP+TN+FN+FP) \tag{13}$$

$$\Pr ecision = TP/(TP+FP) \tag{14}$$

$$F-score = (2 \times TP)/(2 \times TP+FP+FN) \tag{15}$$

$$MCC = \frac{TP \times TN - FP \times FN}{\sqrt{(TP+FP)(TP+FN)(TN+FP)(TN+FN)}} \tag{16}$$

TP, TN, FP, and FN values are the number of true positives, true negatives, false positives, and false negatives, respectively [14].

### 4.1. Classification Results of Subset 1

There are 6000 pieces of 16 × 16 CT patches in Subset-1. Data distribution between classes is equal. 75% of these images were used for training and 25% for testing. Table 2 shows comparatively classification performance pre-trained CNN networks and of the proposed method.

**Table 2.** The classification results for Subset-1

| Methods | Evaluation Metrics (%) | | | | | | | | | |
|---|---|---|---|---|---|---|---|---|---|---|
| | TP | TN | FP | FN | Accuracy | Sensitivity | Specificity | Precision | F1-score | MCC |
| VGG-16 | 716 | 653 | 97 | 34 | 91.27 | 95.47 | 87.07 | 88.07 | 91.62 | 82.83 |
| Resnet-50 | 742 | 682 | 68 | 8 | 94.3 | 98.93 | 90.93 | 91.60 | 95.13 | 90.16 |
| GoogleNet | 631 | 742 | 8 | 119 | 91.53 | 84.13 | 98.93 | 98.75 | 90.86 | 83.99 |
| Proposed Method | 700 | 734 | 16 | 50 | 95.60 | 93.33 | 97.87 | 97.77 | 95.50 | 91.29 |

*4.2. Classification Results of Subset 2*

Subset-2 includes 3000 COVID-19 and 3000 No finding 32 × 32 CT patches. Comparative classification results of Subset-2 are given in Table 3.

**Table 3.** The classification results for Subset-2

| Methods | Evaluation Metrics (%) | | | | | | | | | |
|---|---|---|---|---|---|---|---|---|---|---|
| | TP | TN | FP | FN | Accuracy | Sensitivity | Specificity | Precision | F1-score | MCC |
| VGG-16 | 744 | 710 | 40 | 6 | 96.93 | 99.20 | 94.67 | 94.90 | 97.00 | 93.96 |
| Resnet-50 | 744 | 716 | 34 | 6 | 97.33 | 99.20 | 95.47 | 95.63 | 97.38 | 94.73 |
| GoogleNet | 727 | 741 | 9 | 23 | 97.87 | 96.93 | 98.80 | 98.78 | 97.85 | 95.75 |
| Proposed Method | 742 | 732 | 18 | 8 | 98.27 | 98.93 | 97.60 | 97.63 | 98.28 | 96.54 |

*4.3. Performance Evaluation*

The best performance in Subset-1 showed proposed method with 95.60% as can be seen in Table 2. The highest performance belongs to ResNet-50 models with 98.93% in sensitivity metric. In Specificity and precision metrics, GoogleNet performed best with 98.93% and 98.75% respectively. The proposed method in F1-score and MCC metrics is the most successful among pre-trained CNN structures with 95.50% and 91.29% respectively.

Comparative performance metrics for Subset-2 are given in Table 3. The proposed method stands out with its 98.27% performance in accuracy metric. In the sensitivity metric, VGG-16 and ResNet-50 models show the highest performance with 99.20%. In the precision and F1-score metrics, GoogleNet model achieved 98.80% and 98.78% respectively. The proposed method achieved the highest metric performance in F1-score and MCC metrics with 98.28% and 96.54% respectively. As can be seen in Table 2 and Table 3, there are confusion matrixes obtained with Subset-1 and Subset-2 datasets of the proposed method in Figure 5 and Figure 6.

**Confusion Matrix for Test Data**

|  | Covid-19 | No findings |  |  |
|---|---|---|---|---|
| Covid-19 | 734 | 16 | 97.9% | 2.1% |
| No findings | 50 | 700 | 93.3% | 6.7% |
|  | 93.6% | 97.8% |  |  |
|  | 6.4% | 2.2% |  |  |

**Figure 5.** Confusion Matrix of Proposed Method for Subset-1

Confusion matrix was obtained for proposed method using Subset-1 in Figure 5. When confusion matrix was evaluated in class, COVID-19 class was classified with an accuracy rate of 97.9%. Performance of No findings class was lower than COVID-19. 93.3% accuracy rate was obtained for this class. A classification accuracy of 93.6% was obtained in the analysis of positive class. In negative class, this rate is higher and had a value of 97.8%.

**Confusion Matrix for Test Data**

|  | Covid-19 | No findings |  |  |
|---|---|---|---|---|
| Covid-19 | 732 | 18 | 97.6% | 2.4% |
| No findings | 8 | 742 | 98.9% | 1.1% |
|  | 98.9% | 97.6% |  |  |
|  | 1.1% | 2.4% |  |  |

**Figure 6.** Confusion Matrix of Proposed Method for Subset-2

Subset-2 was used in the training and testing process for the proposed method. In Figure 6, a confusion matrix was obtained for test data. In class analysis, 97.6% accuracy rate of COVID-19 class was obtained. Performance was increased compared to Subset-1 in the No findings class. Accuracy rate was 98.9% for this class. In the positive and negative class evaluation, a classification accuracy of 98.9% and 97.6% was obtained respectively.

## 5. DISCUSSION and CONCLUSION

The first case of COVID-19 was found in the Wuhan region of China. COVID-19 is an epidemic disease and threatens world health system and economy. COVID-19 virus behaves similarly to other pandemic viruses. This makes it difficult to detect COVID-19 cases quickly. Therefore, COVID-19 is a candidate for a global epidemic. Radiological imaging techniques are used for a more accurate diagnosis in the detection of COVID-19. Therefore, it is possible to obtain more detailed information about COVID-19 using CT imaging techniques. When CT images are examined, shadows come to the fore in the regions where COVID-19 is located. At the same time, a spread is observed from the outside to the inner parts.

Obtained images with different CT devices were used in the study. There were different levels of grey level in the images. Different characteristics of CT devices caused it. This complicates the analysis of the images. In the study, deep features were obtained by using pre-trained CNN networks. Then, deep features were fused and ranked. The data set was generated by taking random patches on CT images. Pre-trained CNN networks were trained using the transfer learning method in the Subset-1 and Subset-2 datasets. With the proposed method, 95.60% accuracy, 95.60% sensitivity, 93.33% specificity, 97.87% precision, 97.77% F1-score and 91.29% MCC metric performance were obtained for Subset-1. In Subset-2, proposed method showed 98.27% accuracy, 98.93% sensitivity, 97.60% specificity, 97.63% precision, 98.28% F1-score and 96.54% MCC metric performance.

Most of the studies on COVID-19 are medical studies. Classification and segmentation of COVID-19 images can be performed in the literature. Within the scope of the study, data augmentation techniques can be used to increase the classification performance of COVID-19 images.